\documentclass[11pt]{article}

\usepackage{times}
\usepackage{fullpage}
\usepackage{latexsym}

\def\01{\{0,1\}}

\newcommand{\st}[1]{|#1\rangle} 
 
\newcommand{\inp}[2]{\langle{#1}|{#2}\rangle} % inproduct, < | >
\newcommand{\DISJ}{\mbox{\rm DISJ}} 
\newcommand{\EQ}{\mbox{\rm EQ}} 
\newcommand{\OR}{\mbox{\rm OR}} 
\newcommand{\NOR}{\mbox{\rm NOR}} 
\newcommand{\IP}{\mbox{\rm IP}} 
\newcommand{\so}{{\rm bs0}} 
\newcommand{\eps}{\varepsilon} 

\newtheorem{theorem}{Theorem}
\newtheorem{lemma}{Lemma}
\newtheorem{proposition}{Proposition}
\newtheorem{corollary}{Corollary}

\newenvironment{proof}
{\noindent {\bf Proof }}
{{\hfill $\Box$}\\
 \smallskip}

\begin{document}

\title{Communication Complexity Lower Bounds by Polynomials}
\author{Harry Buhrman\thanks{CWI, P.O.~Box 94709, Amsterdam,
The Netherlands. E-mail: {\tt buhrman@cwi.nl}.
Partially supported by the EU fifth framework project QAIP, IST--1999--11234.}
\and
Ronald de Wolf\thanks{CWI and University of Amsterdam (ILLC). E-mail: {\tt rdewolf@cwi.nl}. 
Partially supported by the EU fifth framework project QAIP, IST--1999--11234.}
}
\date{}
\maketitle

\begin{abstract}
The quantum version of communication complexity allows the two communicating
parties to exchange qubits and/or to make use of prior entanglement 
(shared EPR-pairs).
Some lower bound techniques are available for qubit communication
complexity~\cite{yao:qcircuit,kremer:thesis,cdnt:ip,astv:qsampling}, 
but except for the inner product function~\cite{cdnt:ip}, 
no bounds are known for the model with unlimited prior entanglement.
We show that the ``log rank'' lower bound extends to the strongest model 
(qubit communication $+$ unlimited prior entanglement). 
By relating the rank of the communication matrix 
to properties of polynomials, we are able to derive
some strong bounds for exact protocols. In particular, we prove 
both the ``log-rank conjecture'' and 
the polynomial equivalence of quantum and 
classical communication complexity for various classes of functions.
We also derive some weaker bounds for bounded-error  quantum protocols.
\end{abstract}

\section{Introduction and Statement of Results}

Communication complexity deals with the following kind of problem.
There are two separated parties, usually called Alice and Bob.
Alice receives some input $x\in X$, Bob receives some $y\in Y$,
and together they want to compute some function $f(x,y)$ which
depends on both $x$ and $y$. Alice and Bob are allowed infinite
computational power, but communication between them is expensive
and has to be minimized. How many bits do Alice and Bob have to
exchange in the worst-case in order to be able to compute $f(x,y)$?
This model was introduced by Yao~\cite{yao:distributive} and
has been studied extensively, both for its applications
(like lower bounds on VLSI and circuits) and for its own sake.
We refer to~\cite{kushilevitz&nisan:cc,hromkovic:cc} for definitions and results.

An interesting variant of the above is {\em quantum} communication complexity: 
suppose that Alice and Bob each have a quantum computer at their disposal and are 
allowed to exchange quantum bits (qubits) and/or can make use of the quantum
correlations given by pre-shared EPR-pairs (these
are entangled 2-qubit states $\frac{1}{\sqrt{2}}(\st{00}+\st{11})$
of which Alice has the first qubit and Bob the second) --- can they do 
with fewer communication than in the classical case?
The answer is yes.
Quantum communication complexity was first considered by Yao~\cite{yao:qcircuit} 
and the first example where quantum beats classical communication complexity
was given in~\cite{cleve&buhrman:subs}.
Bigger (even exponential) gaps have been shown since
\cite{BuhrmanCleveWigderson98,astv:qsampling,raz:qcc}.

The question arises how big the gaps
between quantum and classical can be for various (classes of) functions.
In order to answer this, we need to exhibit limits on 
the power of quantum communication complexity, i.e.~establish lower bounds
--- few of which are known currently.
The main purpose of this paper is to develop tools for proving lower
bounds on quantum communication protocols.
We present some new lower bounds for the case where $f$ is a 
total Boolean function. Most of our bounds apply only to exact quantum protocols,
which always output the correct answer. However, we also have some extensions
of our techniques to the case of bounded-error quantum protocols.

\subsection{Lower bounds for exact protocols}

Let $D(f)$ denote the classical deterministic communication complexity of $f$,
$Q(f)$ the qubit communication complexity, and $Q^*(f)$ the qubit communication
required if Alice and Bob can also make use of an unlimited supply of 
pre-shared EPR-pairs.
Clearly $Q^*(f)\leq Q(f)\leq D(f)$.
Ultimately, we would like to show that $Q^*(f)$ and $D(f)$ are 
polynomially related for all total functions $f$ (as are their
query complexity counterparts~\cite{bbcmw:polynomials}).
This requires stronger lower bound tools than we have at present.
Some lower bound methods are available for 
$Q(f)$~\cite{yao:qcircuit,kremer:thesis,cdnt:ip,astv:qsampling},
but the only lower bound known for $Q^*(f)$ is for the inner product function
\cite{cdnt:ip}. A strong and well known lower bound for $D(f)$
is given by the logarithm of the rank of the communication matrix
for $f$~\cite{mehlhorn&schmidt:lasvegas}.
As first noted in~\cite{BuhrmanCleveWigderson98},
techniques of~\cite{yao:qcircuit,kremer:thesis}
imply that an $\Omega(\log rank(f))$-bound also holds for $Q(f)$.
Our first result is to extend this bound to $Q^*(f)$ and to derive the optimal constant:
\begin{equation}\label{eqnlogrank}
Q^*(f)\geq\frac{\log rank(f)}{2}.
\end{equation}
This implies $n/2$ lower bounds for the $Q^*$-complexity of the
equality and disjointness problems, for which no good bounds were known before.
This $n/2$ is tight up to 1 bit, since Alice can send her 
$n$-bit input to Bob with $n/2$ qubits and $n/2$ EPR-pairs 
using superdense coding~\cite{superdense}.
Our corresponding lower bound also
provides a new proof of {\em optimality} of superdense coding.
In fact, the same $n/2$ bound holds for almost all functions.
Furthermore,  proof of the well-known ``log-rank conjecture''
($D(f)\leq (\log rank(f))^k$ for some $k$) would now imply our 
desired polynomial equivalence between $D(f)$ and $Q^*(f)$
(as already noted for $D(f)$ and $Q(f)$ in~\cite{astv:qsampling}).
However, this conjecture is a long standing open question which is probably
hard to solve in full generality.

Secondly, in order to get an algebraic handle on $rank(f)$,
we relate it to a property of polynomials.
It is well known that every total Boolean function 
$g:\01^n\rightarrow\01$ has a unique representation as
a multilinear polynomial in its $n$ variables. 
For the case where Alice and Bob's function has the form $f(x,y)=g(x\wedge y)$, 
we show that $rank(f)$ equals the number of monomials $mon(g)$ of the polynomial 
that represents $g$ ($rank(f)\leq mon(g)$ was shown in~\cite{nisan&wigderson:rank}).
This number of monomials is often easy to count and allows to determine $rank(f)$.
The functions $f(x,y)=g(x\wedge y)$ form an important class which includes
inner product, disjointness, and the functions which give the biggest gaps
known between $D(f)$ and $\log rank(f)$~\cite{nisan&wigderson:rank}
(similar techniques work for $f(x,y)=g(x\vee y)$ or $g(x\oplus y)$).

We use this to show that $Q^*(f)\in\Theta(D(f))$ if $g$ is symmetric.
In this case we also show that $D(f)$ is close to the classical randomized complexity.
Furthermore, $Q^*(f)\leq D(f)\in O(Q^*(f)^2)$ if $g$ is monotone.
For the latter result we rederive a result of Lov\'asz and
Saks~\cite{lovasz&saks:cc} using our tools.

\subsection{Lower bounds for bounded-error protocols}

For the case of bounded-error quantum communication protocols, very few 
lower bounds are currently known (exceptions are inner 
product~\cite{cdnt:ip} and the general discrepancy bound~\cite{kremer:thesis}).
In particular, no good lower bounds are known for the disjointness problem. 
The best known upper bound for this is $O(\sqrt{n}\log n)$ 
qubits~\cite{BuhrmanCleveWigderson98}, contrasting with linear 
classical randomized complexity~\cite{ks:disj,razborov:disj}.
Since disjointness is a co-NP-complete communication problem~\cite{bfs:classes}, 
a good lower bound for this problem would imply lower bounds for all 
NP-hard communication problems.

In order to attack this problem, we make an effort to extend the above 
polynomial-based approach to bounded-error protocols.
We consider the approximate rank $\widetilde{rank}(f)$, and
show the bound $Q_2(f)\geq(\log\widetilde{rank}(f))/2$ for 2-sided
bounded-error qubit protocols 
(again using techniques from~\cite{yao:qcircuit,kremer:thesis}).
Unfortunately, lower bounds on $\widetilde{rank}(f)$ are much harder 
to obtain than for $rank(f)$.
If we could prove for the case $f(x,y)=g(x\wedge y)$ 
that $\widetilde{rank}(f)$ roughly equals the number of monomials 
$\widetilde{mon}(g)$ of an approximating polynomial for $g$,
then a $\sqrt{n}$ lower bound would follow for disjointness,
because we show that this requires at least $2^{\sqrt{n}}$ monomials to approximate.
Since we prove that the quantities $rank(f)$ and $mon(g)$ are in fact equal 
in the exact case, this gives some hope for a similar result
$\widetilde{rank}(f)\approx\widetilde{mon}(g)$ in the approximating case, 
and hence for resolving the complexity of disjointness.

The specific bounds that we actually were able to {\em prove} 
for disjointness are more limited at this point:
$Q_2^*(\DISJ)\in\Omega(\log n)$ for the general case 
(by an extension of techniques of~\cite{cdnt:ip}; the $\log n$ bound 
without entanglement was already known~\cite{astv:qsampling}),
$Q_2^*(\DISJ)\in\Omega(n)$ for 1-round protocols 
(using a result of~\cite{nayak:qfa}), 
and $Q_2(\DISJ)\in\Omega(n)$ if the error probability has to be $<2^{-n}$.

Below we sum up the main results, contrasting the exact and bounded-error case.
\begin{itemize}
\item We show that $Q^*(f)\geq\log rank(f)/2$ for exact protocols 
with unlimited prior EPR-pairs and $Q_2(f)\geq\log\widetilde{rank}(f)/2$
for qubit protocols without prior EPR-pairs.
\item If $f(x,y)=g(x\wedge y)$ for some Boolean function $g$, then $rank(f)=mon(g)$.
An analogous result $\widetilde{rank}(f)\approx\widetilde{mon}(g)$
for the approximate case is open.
\item A polynomial for disjointness, $\DISJ(x,y)=\NOR(x\wedge y)$,
requires $2^n$ monomials in the exact case (implying $Q^*(\DISJ)\geq n/2$), 
and roughly $2^{\sqrt{n}}$ monomials in the approximate case.
\end{itemize}

\section{Preliminaries}

We use $|x|$ to denote the Hamming weight (number of 1s) of $x\in\01^n$,
$x_i$ for the $i$th bit of $x$ ($x_0=0$),
and $e_i$ for the string whose only 1 occurs at position $i$.
If $x,y\in\01^n$, we use $x\wedge y\in\01^n$ for the string obtained by
bitwise ANDing $x$ and $y$, and similarly $x\vee y$.
Let $g:\01^n\rightarrow\01$ be a Boolean function.
We call $g$ {\em symmetric} if $g(x)$ only depends on $|x|$,
and {\em monotone} %(increasing) 
if $g$ cannot decrease if we set more variables to 1.
It is well known that each $g:\01^n\rightarrow R$ has a unique representation
as a multilinear polynomial $g(x)=\sum_{S\subseteq\{1,\ldots,n\}} a_SX_S$,
where $X_S$ is the product of the variables in $S$ and $a_S$
is a real number. The term $a_SX_S$ is called a {\em monomial} of $g$
and $mon(g)$ denotes the number of non-zero monomials of $g$.
A polynomial $p$ {\em approximates} $g$ if $|g(x)-p(x)|\leq 1/3$ for all
$x\in\01^n$.
We use $\widetilde{mon}(g)$ for the minimal number of monomials
among all polynomials which approximate $g$.
The {\em degree} of a monomial is the number of its variables,
and the degree of a polynomial is the largest degree of its monomials.

Let $X$ and $Y$ be finite sets (usually $X=Y=\01^n$)
and $f:X\times Y\rightarrow\01$ be a Boolean function. 
For example, {\em equality} has $\EQ(x,y)=1$ iff $x=y$, 
{\em disjointness} has $\DISJ(x,y)=1$ iff $|x\wedge y|=0$
(equivalently, $\DISJ(x,y)=\NOR(x\wedge y)$), and 
{\em inner product} has $\IP(x,y)=1$ iff $|x\wedge y|$ is odd.
$M_f$ denotes the $|X|\times|Y|$ Boolean matrix whose $x,y$
entry is $f(x,y)$, and $rank(f)$ denotes the rank of $M_f$ over the reals.
A {\em rectangle} is a subset $R=S\times T\subseteq X\times Y$ 
of the domain of $f$.
A {\em 1-cover} for $f$ is a set of rectangles which covers all and only 
1s in $M_f$. $C^1(f)$ denotes the minimal size of a 1-cover for $f$.
For $m\geq 1$, we use $f^{\wedge m}$ to denote the Boolean function which
is the AND of $m$ independent instances of $f$.
That is, $f^{\wedge m}:X^m\times Y^m\rightarrow\01$ and
$f^{\wedge m}(x_1,\ldots,x_m,y_1,\ldots,y_m)=f(x_1,y_1)\wedge
f(x_2,y_2)\wedge\ldots\wedge f(x_m,y_m)$.
Note that $M_{f^{\wedge 2}}$ is the Kronecker product 
$M_f\otimes M_f$ and hence $rank(f^{\wedge m})=rank(f)^m$.

Alice and Bob want to compute some $f:X\times Y\rightarrow\01$.
After the protocol they should both know $f(x,y)$.
Their system has three parts: Alice's part, the 1-qubit channel, and Bob's
part. For definitions of quantum states and operations, 
we refer to~\cite{berthiaume:qc,cleve:intro}.
In the initial state, Alice and Bob share $k$ EPR-pairs and all other qubits
are zero. For simplicity we assume Alice and Bob send 1 qubit in turn, 
and at the end the output-bit of the protocol is put on the channel.
The assumption that 1 qubit is sent per round can be replaced by a fixed
number of qubits $q_i$ for the $i$th round. However, in order to be able 
to run a quantum protocol on a superposition of inputs, it is important that 
the number of qubits sent in the $i$th round is independent of the input $(x,y)$.
An $\ell$-qubit protocol is described by unitary transformations 
$U_1(x),U_2(y),U_3(x),U_4(y),\ldots,U_\ell(x/y)$.
First Alice applies $U_1(x)$ to her part and the channel,
then Bob applies $U_2(y)$ to his part and the channel, etc.  

$Q(f)$ denotes the (worst-case) cost of an optimal qubit protocol that computes
$f$ exactly without prior entanglement,
$C^*(f)$ denotes the cost of a protocol that communicates classical bits but can
make use of an unlimited (but finite) number of shared EPR-pairs,
and $Q^*(f)$ is the cost of a qubit protocol that can use shared EPR-pairs.
$Q_c(f)$ denotes the cost of a {\em clean} qubit protocol without prior
entanglement, i.e.~a protocol that starts with $\st{0}\st{0}\st{0}$ and
ends with $\st{0}\st{f(x,y)}\st{0}$.
We add the superscript ``1 round'' for 1-round protocols, 
where Alice sends a message to Bob and Bob then sends the output bit.
Some simple relations that hold between these measures
are $Q^*(f)\leq Q(f)\leq D(f)\leq D^{1 round}(f)$, 
$Q(f)\leq Q_c(f)\leq 2Q(f)$ and $Q^*(f)\leq C^*(f)\leq 2Q^*(f)$~\cite{teleporting}.
For bounded-error protocols we analogously define $Q_2(f)$, $Q_2^*(f)$,
$C^*_2(f)$ for quantum protocols that give the correct answer with probability 
at least $2/3$ on every input.
We use $R^{pub}_2(f)$ for the classical bounded-error complexity in the
public-coin model~\cite{kushilevitz&nisan:cc}.

\section{Log-Rank Lower Bound}\label{seclogrank}

As first noted in~\cite{BuhrmanCleveWigderson98,astv:qsampling}, techniques 
from~\cite{yao:qcircuit,kremer:thesis} imply $Q(f)\in\Omega(\log rank(f))$. 
For completeness we prove the following $\log rank(f)$ bound for clean quantum
protocols in Appendix~\ref{applogrankclean}. 
This implies $Q(f)\geq\log rank(f)/2$.
We then extend this to the case where Alice and Bob share prior entanglement:%
\footnote{During discussions we had with Michael Nielsen  in Cambridge in 
the summer of 1999, it appeared that an equivalent result can be derived from
results about {\em Schmidt numbers} in~\cite[Section~6.4.2]{nielsen:thesis}.}

\begin{theorem}\label{thlogrankclean}
$\displaystyle Q_c(f)\geq\log rank(f)+1$.
\end{theorem}

\begin{theorem}\label{thlogrank+ent}
$\displaystyle Q^*(f)\geq\frac{\log rank(f)}{2}$.
\end{theorem}

\begin{proof}
Suppose we have some exact protocol for $f$ that uses $\ell$ qubits 
of communication and $k$ prior EPR-pairs. We will build a clean qubit 
protocol without prior entanglement for $f^{\wedge m}$.
First Alice makes $k$ EPR-pairs and sends one half of each pair to Bob
(at a cost of $k$ qubits of communication). Now they run the protocol
to compute the first instance of $f$ ($\ell$ qubits of communication). 
Alice copies the answer to a safe place which we will call the `answer bit'
and they reverse the protocol (again $\ell$ qubits of communication). 
This gives them back the $k$ EPR-pairs, which they can reuse.  
Now they compute the second instance of $f$, Alice ANDs the answer into 
the answer bit (which can be done cleanly), and they reverse the protocol, etc.
After all $m$ instances of $f$ have been computed, Alice and Bob
have the answer $f^{\wedge m}(x,y)$ left and the $k$ EPR-pairs, 
which they uncompute using another $k$ qubits of communication.

This gives a clean protocol for $f^{\wedge m}$ that uses $2m\ell+2k$ qubits 
and no prior entanglement. By Theorem~\ref{thlogrankclean}:
$$
2m\ell+2k \geq Q_c(f^{\wedge m}) \geq\log rank(f^{\wedge m})+1 = m\log rank(f)+1,
$$
hence
$$
\ell\geq\frac{\log rank(f)}{2}-\frac{2k-1}{2m}.
$$
Since this must hold for every $m>0$, the theorem follows.
\end{proof}

We can derive a stronger bound for $C^*(f)$:

\begin{theorem}\label{thlogrankc*}
$\displaystyle C^*(f)\geq \log rank(f)$.
\end{theorem}

\begin{proof}
Since a qubit and an EPR-pair can be used to send 2 classical bits~\cite{superdense},
we can devise a qubit protocol for $f\wedge f$ using $C^*(f)$ qubits
(compute the two copies of $f$ in parallel using the classical bit protocol).
Hence by the previous theorem
$C^*(f)\geq Q^*(f\wedge f)\geq (\log rank(f\wedge f))/2=\log rank(f)$.
\end{proof}

%\subsection{Some consequences}
Below we draw some consequences from these log-rank lower bounds.
Firstly, $M_{\rm EQ}$ is the identity matrix, so $rank(\EQ)=2^n$.
This gives the bounds $Q^*(\EQ)\geq n/2$, $C^*(\EQ)\geq n$
(in contrast, $Q_2(\EQ)\in\Theta(\log n)$ and $C^*_2(\EQ)\in O(1)$).
The disjointness function on $n$ bits is the AND of $n$ disjointnesses on 
1 bit (which have rank 2 each), so $rank(\DISJ)=2^n$. 
The complement of the inner product function has $rank(f)=2^n$.
Thus we have the following strong lower bounds, all tight up to 1 bit:%
\footnote{The same bounds for \IP\ are also given in~\cite{cdnt:ip}.
The bounds for \EQ\ and \DISJ\ are new, and can also be shown to hold
for {\em zero-error} quantum protocols.}

\begin{corollary}
$Q^*(\EQ), Q^*(\DISJ), Q^*(\IP)\geq n/2$ and $C^*(\EQ), C^*(\DISJ), C^*(\IP)\geq n$.
\end{corollary}

%We can also get tight bounds for the $\DISJ_k$ function, 
%which is the special case of disjointness where the inputs of Alice
%and Bob have weight $k$ (so $|X|=|Y|={n\choose k}$).
%Since Alice can just send the index of her input to Bob, 
%we have trivial upper bounds 
%$C^*(DISJ_k)\leq D(\DISJ_k)\leq\log{n\choose k}+1$ 
%and $Q^*(\DISJ_k)\leq\log{n\choose k}/2+1$.
%On the other hand, the communication matrix for $\DISJ_k$ 
%has full rank (see~\cite[Example~2.12]{kushilevitz&nisan:cc}),
%which gives the lower bounds $C^*(\DISJ_k)\geq\log{n\choose k}$ 
%and $Q^*(\DISJ_k)\geq\log{n\choose k}/2$.
%For instance, for $k=\sqrt{n}$ we get $Q^*(\DISJ_k)\in\Theta(\sqrt{n}\log n)$.

Koml\'os~\cite{komlos:det01} has shown that the fraction of $m\times m$ 
Boolean matrices that have determinant 0 goes to 0 as $m\rightarrow\infty$. 
%hromkovic, Ex 2.1.5.22, &p145
Hence almost all $2^n\times 2^n$ Boolean matrices have full rank $2^n$, 
which implies that almost all functions have maximal quantum communication 
complexity:

\begin{corollary}
Almost all $f:\01^n\times\01^n\rightarrow\01$ have 
$Q^*(f)\geq n/2$ and $C^*(f)\geq n$.
\end{corollary}

%Call a function {\em non-redundant} if all rows of $M_f$ are different
%(if the $i$th and $j$th row of $M_f$ are the same, then Alice's $i$th
%and $j$th input can be identified, thereby reducing her domain).
%A weak lower bound for {\em all} non-redundant $f$ is:
%
%\begin{corollary}
%If $f$ is non-redundant, then $Q^*(f)\geq (\log n)/2$ and $C^*(f)\geq \log n$.
%\end{corollary}
%
%\begin{proof}
%It suffices to show that $rank(f)\geq n$.
%Suppose $rank(f)=r<n$. Then there are $r$ columns $c_1,\ldots,c_r$ in $M_f$
%which span the column space of $M_f$. Let $A$ be the $2^n\times r$
%matrix that has these $c_i$ as columns. Let $B$ be the $r\times 2^n$
%matrix whose $i$th column is formed by the $r$ coefficients of the $i$th column of
%$M_f$ when written out as a linear combination of $c_1,\ldots,c_r$. Then $M_f=AB$.
%Because the $2^n$ rows of $A$ are Boolean and $r<n$, 
%there are two rows in $A$ that are the same, say the $i$th and $j$th.
%But then the $i$th and $j$th row of $M_f$ are also the same, contradiction.
%\end{proof}

We say $f$ satisfies the {\em quantum direct sum property} if computing
$m$ independent copies of $f$ (without prior entanglement) takes $mQ(f)$
qubits of communication in the worst case. 
(We have no example of an $f$ without this property.)
Using the same technique as before, we can prove an equivalence
between the qubit models with and without prior entanglement for such $f$:

\begin{corollary}
If $f$ satisfies the quantum direct sum property, then
$Q^*(f)\leq Q(f)\leq 2Q^*(f)$.
\end{corollary}

\begin{proof}
$Q^*(f)\leq Q(f)$ is obvious. 
Using the techniques of Theorem~\ref{thlogrank+ent} we have 
$mQ(f)\leq 2mQ^*(f)+k$, for all $m$ and some fixed $k$, 
hence $Q(f)\leq 2Q^*(f)$.
\end{proof}

Finally, because of Theorem~\ref{thlogrank+ent},
the well-known ``log-rank conjecture'' now implies the polynomial
equivalence of deterministic classical communication complexity and 
exact quantum communication complexity (with or without prior entanglement)
for all total $f$:

\begin{corollary}
If $D(f)\in O((\log rank(f))^k)$, then $Q^*(f)\leq Q(f)\leq D(f)\in
O(Q^*(f)^k)$ for all $f$.
\end{corollary}

%For example, Faigle, Schrader and Tur\'an~\cite{fst:ccinterval}
%prove that if $f$ is a ``generalized interval order'', then
%$D(f)$ equals $\log rank(f)$ up to 1 bit.
%Hence $C^*(f)$ equals $D(f)$ up to 1 bit, and $Q^*(f)$ equals
%$D(f)$ up to a factor of 2 for such $f$.

\section{A Lower Bound Technique via Polynomials} 

\subsection{Decompositions and polynomials}

The previous section showed that lower bounds on $rank(f)$ imply
lower bounds on $Q^*(f)$. In this section we relate $rank(f)$ to
the number of monomials of a polynomial for $f$ 
and use this to prove lower bounds for some classes of functions.

We define the {\em decomposition number} $m(f)$ of some function 
$f:\01^n\times\01^n\rightarrow R$ as the minimum $m$ such that there 
exist functions $a_1(x),\ldots,a_m(x)$ and $b_1(y),\ldots,b_m(y)$ 
(from $R^n$ to $R$) for which $f(x,y)=\sum_{i=1}^m a_i(x)b_i(y)$ for all $x,y$.
We say that $f$ can be {\em decomposed} into the $m$ functions $a_ib_i$.
Without loss of generality, the functions $a_i,b_i$ may be assumed
to be multilinear polynomials.
It turns out that the decomposition number equals the rank:%
\footnote{The first part of the proof employs a technique of Nisan and 
Wigderson~\cite{nisan&wigderson:rank}. 
They used this to prove $\log rank(f)\in O(n^{\log_3 2})$
for a specific $f$. Our Corollary~\ref{corm=mon} below implies
that this is tight: $\log rank(f)\in\Theta(n^{\log_3 2})$ for their $f$.}

\begin{lemma}\label{lemrank=m}
$rank(f)=m(f)$.
\end{lemma}

\begin{proof}

{$\bf rank(f)\leq m(f)$:}
Let $f(x,y)=\sum_{i=1}^m a_i(x)b_i(y)$,
$M_i$ be the matrix defined by $M_i(x,y)=a_i(x)b_i(y)$,
$r_i$ be the row vector whose $y$th entry is $b_i(y)$.
Note that the $x$th row of $M_i$ is $a_i(x)$ times $r_i$.
Thus {\em all} rows of $M_i$ are scalar multiples of each other, 
hence $M_i$ has rank 1.
Since $rank(A+B)\leq rank(A)+rank(B)$ and $M_f=\sum_{i=1}^{m(f)}M_i$,
we have $rank(f)=rank(M_f)\leq\sum_{i=1}^{m(f)} rank(M_i)= m(f)$.

{$\bf m(f)\leq rank(f)$:}
Suppose $rank(f)=r$. Then there are $r$ columns $c_1,\ldots,c_r$ in $M_f$
which span the column space of $M_f$. Let $A$ be the $2^n\times r$
matrix that has these $c_i$ as columns. Let $B$ be the $r\times 2^n$
matrix whose $i$th column is formed by the $r$ coefficients of the $i$th 
column of $M_f$ when written out as a linear combination of $c_1,\ldots,c_r$. 
Then $M_f=AB$, hence
$f(x,y)=M_f(x,y)=\sum_{i=1}^r A_{xi}B_{iy}.$
Defining functions $a_i,b_i$ by $a_i(x)=A_{xi}$ and $b_i(y)=B_{iy}$,
we have $m(f)\leq rank(f)$.
\end{proof}

Combined with Theorems~\ref{thlogrank+ent} and~\ref{thlogrankc*} we obtain

\begin{corollary}
$\displaystyle Q^*(f)\geq\frac{\log m(f)}{2}$ and
$\displaystyle C^*(f)\geq\log m(f)$.
\end{corollary}

Accordingly, for lower bounds on quantum communication complexity it is
important to be able to determine the decomposition number $m(f)$. 
Often this is hard.
It is much easier to determine the number of monomials $mon(f)$ of $f$ 
(which upper bounds $m(f)$). Below we show that in the special case where 
$f(x,y)=g(x\wedge y)$, these two numbers are the same.%
\footnote{After learning about this result, Mario Szegedy (personal
communication) came up with an alternative proof of this, using Fourier transforms.}

Below, a monomial is called {\em even} if it contains $x_i$ iff it contains $y_i$,
for example $2x_1x_3y_1y_3$ is even and $x_1x_3y_1$ is not.
A polynomial is even if each of its monomials is even.

\begin{lemma}
If $p:\01^n\times\01^n\rightarrow R$ is an even polynomial with $k$
monomials, then $m(p)=k$.
\end{lemma}

\begin{proof}
Clearly $m(p)\leq k$. To prove the converse, 
consider $\DISJ(x,y)=\Pi_{i=1}^n(1-x_iy_i)$, the unique polynomial 
for the disjointness function. Note that this polynomial contains
all and only even monomials (with coefficients $\pm 1$).
Since \DISJ\ has rank $2^n$, it follows from Lemma~\ref{lemrank=m}
that \DISJ\ cannot be decomposed in fewer then $2^n$ terms.
We will show how a decomposition of $p$ with $m(p)<k$ would give
rise to a decomposition of \DISJ\ with fewer than $2^n$ terms.
Suppose we can write
$$
p(x,y)=\sum_{i=1}^{m(p)}a_i(x)b_i(y).
$$
Let $aX_SY_S$ be some even monomial in $p$ and suppose 
the monomial $X_SY_S$ in \DISJ\ has coefficient $c=\pm1$. 
Now whenever $bX_S$ occurs in some $a_i$, replace that $bX_S$ by $(cb/a)X_S$.
Using the fact that $p$ contains only even monomials, 
it is not hard to see that the new polynomial obtained in this way is 
the same as $p$, except that the monomial $aX_SY_S$ is replaced by $cX_SY_S$.

Doing this sequentially for all monomials in $p$,
we end up with a polynomial $p'$ (with $k$ monomials and 
$m(p')\leq m(p)$) which is a subpolynomial of \DISJ, in the sense
that each monomial in $p'$ also occurs with the same coefficient in \DISJ.
Notice that by adding all $2^n-k$ missing \DISJ-monomials to $p'$, 
we obtain a decomposition of \DISJ\ with $m(p')+2^n-k$ terms.
But any such decomposition needs at least $2^n$ terms,
hence $m(p')+2^n-k\geq 2^n$, which implies $k\leq m(p')\leq m(p)$.
\end{proof}

If $f(x,y)=g(x\wedge y)$ for some Boolean function $g$, 
then the polynomial that represents $f$ is just the polynomial of $g$ 
with the $i$th variable replaced by $x_iy_i$.
Hence such a polynomial is even, and we obtain:

\begin{corollary}\label{corm=mon}
If $g:\01^n\rightarrow\01$ and $f(x,y)=g(x\wedge y)$,
then $mon(g)=mon(f)=m(f)=rank(f)$.
\end{corollary}

This gives a strong tool for lower bounding (quantum and classical)
communication complexity whenever $f$ is of the form $f(x,y)=g(x\wedge y)$:
$\log mon(g)\leq C^*(f)\leq D(f)$.
Below we give some applications.

\subsection{Symmetric functions}

As a first application we show that $D(f)$ and $Q^*(f)$ are linearly related
if $f(x,y)=g(x\wedge y)$ and $g$ is symmetric (this follows from 
Corollary~\ref{corDCrelated} below).
Furthermore, we show that the classical randomized public-coin complexity 
$R_2^{pub}(f)$ can be at most a $\log n$-factor less than $D(f)$ for such $f$
(Theorem~\ref{thDRsymm}).
We will assume without loss of generality that $g(\vec{0})=0$, so the
polynomial representing $g$ does not have the constant-1 monomial.

\begin{lemma}\label{lemsymupper}
If $g$ is a symmetric function whose lowest-weight 1-input has Hamming weight $t>0$ 
and $f(x,y)=g(x\wedge y)$, then 
$D^{1 round}(f)=\log\left(\sum_{i=t}^n{n\choose i}+1\right)+1$.
\end{lemma}

\begin{proof}
It is known (and easy to see) that $D^{1 round}(f)=\log r+1$,
where $r$ is the number of different rows of $M_f$ (this equals 
the number of different columns in our case, because $f(x,y)=f(y,x)$).
We count $r$.
Firstly, if $|x|<t$ then the $x$-row contains only zeroes.
Secondly, if $x\neq x'$ and both $|x|\geq t$ and $|x'|\geq t$
then it is easy to see that there exists a $y$ such that $|x\wedge y|=t$
and $|x'\wedge y|<t$ (or vice versa), hence $f(x,y)\neq f(x',y)$ so
the $x$-row and $x'$-row are different.
Accordingly, $r$ equals the number of different $x$ with $|x|\geq t$, 
$+1$ for the 0-row, which gives the lemma.
\end{proof}

\begin{lemma}\label{lemsymmon}
If $g$ is a symmetric function whose lowest-weight 1-input has weight $t>0$, then\\
$(1-o(1))\log\left(\sum_{i=t}^n{n\choose i}\right) \leq \log mon(g)\leq 
\log\left(\sum_{i=t}^n{n\choose i}\right).$
\end{lemma}

\begin{proof}
The upper bound follows
from the fact that $g$ cannot have monomials of degree $<t$.
For the lower bound we distinguish two cases.

{\bf Case 1: $\bf t\leq n/2$.}
It is known that every symmetric $g$ has degree 
$deg(g)=n-O(n^{0.548})$~\cite{gathen&roche:poly}.
That is, an interval $I=[a,n]$ such that $g$ has no monomials of 
any degree $d\in I$ has length at most $O(n^{0.548})$.
This implies that every interval $I=[a,b]$ ($b\geq t$) such that $g$ has 
no monomials of any degree $d\in I$ has length at most $O(n^{0.548})$
(by setting $n-b$ variables to 0, we can reduce to a function on $b$
variables where $I$ occurs ``at the end'').
Since $g$ must have monomials of degree $t\leq n/2$, $g$ must contain 
a monomial of degree $d$ for some $d\in[n/2,n/2+O(n^{0.548})]$.
But because $g$ is symmetric, it must then contain {\em all} ${n\choose d}$ 
monomials of degree $d$.
Hence by Stirling's approximation $mon(g)\geq{n\choose d}\geq 2^{n-O(n^{0.548})}$, 
which implies the lemma.

{\bf Case 2: $\bf t>n/2$.}
It is easy to see that $g$ must contain all ${n\choose t}$ 
monomials of degree $t$.
Now
$$
(n-t+1)mon(g)\geq (n-t+1){n\choose t}\geq\sum_{i=t}^n{n\choose i}.
$$
Hence $\log mon(g)\geq\log\left(\sum_{i=t}^n{n\choose i}\right)
-\log(n-t+1)=(1-o(1))\log\left(\sum_{i=t}^n{n\choose i}\right)$.
\end{proof}

The number $mon(g)$ may be less then $\sum_{i=t}^n{n\choose i}$.
Consider the function
$g(x_1,x_2,x_3)=x_1+x_2+x_3-x_1x_2-x_1x_3-x_2x_3$~\cite{nisan&szegedy:degree}.
Here $mon(g)=6$ but $\sum_{i=1}^3{3\choose i}=7$.
Hence the $1-o(1)$ of Lemma~\ref{lemsymmon} cannot be improved to $1$
in general (it can if $g$ is a threshold function). 

Combining the previous results:

\begin{corollary}\label{corDmon}
If $g$ is a symmetric function whose lowest-weight 1-input 
has weight $t>0$ and $f(x,y)=g(x\wedge y)$, then
$(1-o(1))\log\left(\sum_{i=t}^n{n\choose i}\right) \leq C^*(f)\leq D(f)
\leq D^{1 round}(f) =\log\left(\sum_{i=t}^n{n\choose i}+1\right)+1.$
\end{corollary}

Accordingly, for symmetric $g$ the communication complexity (quantum and 
classical, with or without prior entanglement, 1-round and multi-round) 
equals $\log rank(f)$ up to small constant factors.
In particular:

\begin{corollary}\label{corDCrelated}
If $g$ is symmetric and $f(x,y)=g(x\wedge y)$, 
then $(1-o(1))D(f)\leq C^*(f)\leq D(f)$.
\end{corollary}

We have shown that $Q^*(f)$ and $D(f)$ are equal up to constant
factors whenever $f(x,y)=g(x\wedge y)$ and $g$ is symmetric.
For such $f$, $D(f)$ is also nearly equal to the classical
bounded-error communication complexity $R^{pub}_2(f)$, where we
allow Alice and Bob to share public coin flips.
In order to prove this, we introduce the notion of {\em 0-block sensitivity}
in analogy to the notion of block sensitivity of Nisan~\cite{nisan:pram&dt}.
For input $x\in\01^n$, let $\so_x(g)$ be the maximal number of disjoint sets
$S_1,\ldots,S_b$ of indices of variables, such that for every $i$ we have
(1) all $S_i$-variables have value 0 in $x$ and (2) $g(x)\neq g(x^{S_i})$,
where $x^{S_i}$ is the string obtained from $x$ by setting all $S_i$-variables
to 1. Let $\so(g)=\max_x \so_x(g)$.
We now have:

\begin{lemma}\label{lemsymmonso}
If $g$ is a symmetric function,
then $mon(g)\leq n^{2 \so(g)}$.
\end{lemma}

\begin{proof}
Let $t$ be the smallest number such that $g_t\neq g_{t+1}$, then $\so(g)\geq n-t$.
If $t\leq n/2$ then $\so(g)\geq n/2$, so $mon(g)\leq 2^n\leq n^{2 \so(g)}$.
If $t>n/2$ then $g$ has no monomials of degree $\leq t$, hence
$mon(g)\leq\sum_{i=t+1}^n{n\choose i}
%=\sum_{i=0}^{n-t}{n\choose t}\leq\left(\frac{en}{n-t}\right)^{n-t}
\leq n^{2 \so(g)}.$
\end{proof}

%Now $R^{pub}_2(f)$ cannot be much less than $D(f)$:

\begin{theorem}\label{thDRsymm}
If $g$ is a symmetric function and $f(x,y)=g(x\wedge y)$,
then $D(f)\in O(R^{pub}_2(f)\log n)$.
\end{theorem}

\begin{proof}
By Corollary~\ref{corDmon} we have $D(f)\leq (1+o(1))\log mon(g)$.
Lemma~\ref{lemsymmonso} implies $D(f)\in O(\so(g)\log n)$.
Using Razborov's lower bound technique for disjointness~\cite{razborov:disj}
(see also~\cite[Section~4.6]{kushilevitz&nisan:cc}) we can easily show 
$R^{pub}_2(f)\in\Omega(\so(f))$, which implies the theorem.
\end{proof}

This theorem is tight for the function defined by $g(x)=1$ iff $|x|\geq n-1$.
We have $mon(g)=n+1$, so $\log n\leq D(f)\leq (1+o(1))\log n$.
On the other hand, an $O(1)$ bounded-error public coin protocol
can easily be derived from the well-known $O(1)$-protocol for equality:
Alice tests if $|x|<n-1$, sends a 0 if so and a 1 if not.
In the first case Alice and Bob know that $f(x,y)=0$.
In the second case, we have $f(x,y)=1$ iff $x=y$ or $y=\vec{1}$,
which can be tested with 2 applications of the equality-protocol.
Hence $R^{pub}_2(f)\in O(1)$.

\subsection{Monotone functions}

A second application concerns monotone problems.
Lov\'{a}sz and Saks~\cite{lovasz&saks:cc} prove the log-rank conjecture
for (among others) the following problem, which they call the
{\em union problem for $\bf C$}. Here $\bf C$ is a monotone set system
(i.e.~$(A\in{\bf C}\wedge A\subseteq B)\Rightarrow B\in{\bf C}$)
over some size-$n$ universe. Alice and Bob receive sets $x$ and $y$ (respectively) 
from this universe, and their task is to determine whether $x\cup y\in{\bf C}$.
Identifying sets with their representation as $n$-bit strings, 
this problem can equivalently be viewed as a function $f(x,y)=g(x\vee y)$, 
where $g$ is a monotone increasing Boolean function.
Note that it doesn't really matter whether we take $g$ increasing or decreasing,
nor whether we use $x\vee y$ or $x\wedge y$, as these problems can all be
converted into each other via De Morgan's laws.
Our translation of rank to number of monomials now allows us to  
rederive the Lov\'{a}sz-Saks result without making use of their 
combinatorial lattice theoretical machinery.
We just need the following, slightly modified, result from their paper
(a proof is given in Appendix~\ref{appsakslovasz}):

\begin{theorem}[Lov\'{a}sz and Saks]\label{thnlogrank}
$D(f)\leq (1+\log(C^1(f)+1))(2+\log rank(f))$.
\end{theorem}

\begin{theorem}[Lov\'{a}sz and Saks]
If $g$ is monotone and $f(x,y)=g(x\wedge y)$, then $D(f)\in O((\log rank(f))^2)$.
\end{theorem}

\begin{proof}
Let $M_1,\ldots,M_k$ be all the minimal monomials in $g$.
Each $M_i$ induces a rectangle $R_i=S_i\times T_i$, where 
$S_i=\{x \mid M_i\subseteq x\}$ and $T_i=\{y \mid M_i\subseteq y\}$.
Because $g$ is monotone increasing, $g(z)=1$ iff $z$ makes at least one $M_i$
true. Hence $f(x,y)=1$ iff there is an $i$ such that $(x,y)\in R_i$.
Accordingly, the set of $R_i$ is a 1-cover for $f$ and 
$C^1(f)\leq k\leq mon(g)=rank(f)$ by Corollary~\ref{corm=mon}.
Plugging into Theorem~\ref{thnlogrank} gives the theorem.
\end{proof}

\begin{corollary}
If $g$ is monotone and $f(x,y)=g(x\wedge y)$, then $D(f)\in O(Q^*(f)^2)$.
\end{corollary}

This result can be tightened for the special case of $d$-level AND-OR-trees.
For example, let $g$ be a 2-level AND-of-ORs on $n$ variables with fan-out 
$\sqrt{n}$ and $f(x,y)=g(x\wedge y)$. Then $g$ has
$(2^{\sqrt{n}}-1)^{\sqrt{n}}$ monomials and hence $Q^*(f)\geq n/2$.
In contrast, the zero-error quantum complexity of $f$ is 
$O(n^{3/4}\log n)$~\cite{bcwz:qerror}.

\section{Bounded-Error Protocols}\label{secqccberror}

%\subsection{A polynomial method}

Here we generalize the above approach to bounded-error quantum protocols.
Define the {\em approximate rank} of $f$, $\widetilde{rank}(f)$, 
as the minimum rank among all matrices
$M$ that approximate $M_f$ entry-wise up to $1/3$.
Let the {\em approximate decomposition number}
$\widetilde{m}(f)$ be the minimum $m$ such that there exist 
functions $a_1(x),\ldots,a_m(x)$ and $b_1(y),\ldots,b_m(y)$ 
for which $|f(x,y)-\sum_{i=1}^m a_i(x)b_i(y)|\leq 1/3$ for all $x,y$.
By the same proof as for Lemma~\ref{lemrank=m} we obtain:

\begin{lemma}\label{lemrank=mappr}
$\widetilde{rank}(f)=\widetilde{m}(f)$.
\end{lemma}

By a proof similar to Theorem~\ref{thlogrankclean} 
(again using methods from~\cite{yao:qcircuit,kremer:thesis}, 
see Appendix~\ref{applogrankappr}) we show

\begin{theorem}\label{thlogrankappr}
$\displaystyle Q_2(f)\geq\frac{\log\widetilde{m}(f)}{2}$.
\end{theorem}

Unfortunately, it is much harder to prove bounds on $\widetilde{m}(f)$ than on $m(f)$.% 
\footnote{It is interesting to note that $\overline{\IP}$ (the negation of \IP)
has less than maximal approximate decomposition number. For example
for $n=2$, $m(f)=4$ but $\widetilde{m}(f)=3$.}
In the exact case we have $m(f)=mon(g)$ whenever $f(x,y)=g(x\wedge y)$,
and $mon(g)$ is often easy to determine.
If something similar is true in the approximate case, then we obtain
strong lower bounds on $Q_2(f)$, because our next theorem gives a 
bound on $\widetilde{mon}(g)$ in terms of the 0-block sensitivity
defined in the previous section
(the proof is deferred to Appendix~\ref{appmonbound}).

\begin{theorem}\label{thapprmon}
If $g$ is a Boolean function, then
$\displaystyle\widetilde{mon}(g)\geq 2^{\sqrt{\so(g)/12}}.$
\end{theorem}

In particular, for $\DISJ(x,y)=\NOR(x\wedge y)$
it is easy to see that $\so(\NOR)=n$,  hence $\log\widetilde{mon}(\NOR)\geq\sqrt{n/12}$
(the upper bound  $\log\widetilde{mon}(\NOR)\in O(\sqrt{n}\log n)$
follows from the construction of a degree-$\sqrt{n}$ polynomial
for \OR\ in~\cite{nisan&szegedy:degree}).
Consequently, a proof that the approximate decomposition number $\widetilde{m}(f)$
roughly equals $\widetilde{mon}(g)$ would give $Q_2(\DISJ)\in\Omega(\sqrt{n})$,
nearly matching the $O(\sqrt{n}\log n)$ upper bound of~\cite{BuhrmanCleveWigderson98}.
Since $m(f)=mon(g)$ in the exact case, a result like 
$\widetilde{m}(f)\approx\widetilde{mon}(g)$ might be doable.

We end this section by proving some weaker lower bounds for disjointness.
Firstly, disjointness has a bounded-error protocol with $O(\sqrt{n}\log n)$ 
qubits and $O(\sqrt{n})$ rounds~\cite{BuhrmanCleveWigderson98}, 
but if we restrict to 1-round protocols then a linear lower bound
follows from a result of Nayak~\cite{nayak:qfa}:

\begin{theorem}
$Q_2^{1 round}(\DISJ)\in\Omega(n)$.
\end{theorem}

\begin{proof}
Suppose there exists a 1-round qubit protocol with $m$ qubits:
Alice sends a message $M(x)$ of $m$ qubits to Bob, and Bob then
has sufficient information to establish whether Alice's $x$ and Bob's $y$
are disjoint. 
Note that $M(x)$ is independent of $y$. 
If Bob's input is $y=e_i$, then $\DISJ(x,y)$ is the negation of Alice's $i$th bit. 
But then the message is an $(n,m,2/3)$ quantum random access 
code~\cite{antv:dense}: by choosing input $y=e_i$ and continuing 
the protocol, Bob can extract from $M(x)$ the $i$th bit of Alice 
(with probability $\geq 2/3$), for any $1\leq i\leq n$ of his choice.
For this the lower bound $m\geq (1-H(2/3))n>0.08\ n$ is known~\cite{nayak:qfa}.
\end{proof}

For multi-round quantum protocols for disjointness with bounded error
probability we can only prove a logarithmic lower bound, using a technique 
from~\cite{cdnt:ip} (we omit the proof for reasons of space;
for the model without entanglement, the bound 
$Q_2(\DISJ)\in\Omega(\log n)$ was already shown in~\cite{astv:qsampling}).

\begin{proposition}
$Q_2^*(\DISJ)\in\Omega(\log n)$.
\end{proposition}

%\begin{proofsketch}
%We sketch the proof for a protocol which maps
%$\st{x}\st{y}\rightarrow(-1)^{{\rm DISJ}(x,y)}\st{x}\st{y}$.
%Alice chooses some $i\in\{1,\ldots,n\}$ and starts with $\st{e_i}$, Bob starts with
%$(1/\sqrt{2^n})\sum_y\st{y}$. After running the  protocol, Bob has state
%$$
%\st{\phi_i} = \sum_y\frac{(-1)^{{\rm DISJ}(e_i,y)}}{\sqrt{2^n}}\st{y}
%            = \sum_y\frac{(-1)^{1-y_i}}{\sqrt{2^n}}\st{y}.
%$$
%Note that
%$$
%\inp{\phi_i}{\phi_j}=\frac{1}{2^n}\sum_y(-1)^{y_i+y_j}=\delta_{ij}.
%$$
%Hence the $\st{\phi_i}$ form an orthogonal set, and Bob can determine
%exactly which $\st{\phi_i}$ he has and thus learn $i$. Alice now has 
%transmitted $\log n$ bits to Bob and Holevo's theorem~\cite{holevo} 
%implies that at least $(\log n)/2$ qubits must have been communicated 
%to achieve this.
%A similar analysis works for bounded-error (as in~\cite{cdnt:ip}).
%\end{proofsketch}

Finally, for the case where we want to compute disjointness with
very small error probability, we can prove an $\Omega(n)$ bound.
Here we use the subscript ``$\eps$'' to indicate qubit protocols
(without prior entanglement) whose error probability is $\leq\eps$.
We first give a bound for equality:

\begin{theorem}
If $\eps<2^{-n}$, then $Q_\eps(\EQ)\geq n/2$.
\end{theorem}

\begin{proof}
By Lemma~\ref{lemrank=mappr} and Theorem~\ref{thlogrankappr}, 
it suffices to show that an $\eps$-approximation of 
the $2^n\times 2^n$ identity matrix $I$ requires full rank. 
Suppose that $M$ approximates $I$ entry-wise up to $\eps$ but has rank $<2^n$. 
Then $M$ has some eigenvalue $\lambda=0$.
Ger\u{s}gorin's Disc Theorem (see~\cite[p.31]{horn&johnson:topics}) 
implies that all eigenvalues of $M$ are in the set
$\bigcup_i\{z\mid |z-M_{ii}|\leq R_i\},$
where $R_i=\sum_{j\neq i}|M_{ij}|$.
But if $\lambda=0$ is in this set, then for some $i$
$1-\eps\leq|M_{ii}|=|\lambda-M_{ii}|\leq R_i\leq (2^n-1)\eps,$
hence $\eps\geq 2^{-n}$, contradiction.
\end{proof}

We reduce equality to disjointness.
Let $x,y\in\01^n$.
Define $x'\in\01^{2n}$ by replacing $x_i$ by $x_i\overline{x_i}$ in $x$,
and $y'\in\01^{2n}$ by replacing $y_i$ by $\overline{y_i}y_i$ in $y$.
It is easy to see that $\EQ(x,y)=\DISJ(x',y')$ so we have:

\begin{corollary}
If $\eps<2^{-n}$, then $Q_\eps(\DISJ)\geq n/4$.
\end{corollary}

\section{Open Problems}

To end this paper, we identify three important open questions 
in quantum communication complexity.
First, are $Q^*(f)$ and $D(f)$ polynomially related for {\em all} total $f$,
or at least for all $f$ of the form $f(x,y)=g(x\wedge y)$?
We have proven this for some special cases here 
($g$ symmetric or monotone), but the general question remains open.
There is a close analogy between the quantum communication complexity 
lower bounds presented here, and the quantum query complexity bounds
obtained in~\cite{bbcmw:polynomials}. Let $deg(g)$ and $mon(g)$ be,
respectively, the degree and the number of monomials of the polynomial
that represents $g:\01^n\rightarrow\01$.
In~\cite{bbcmw:polynomials} it was shown that a quantum computer needs 
at least $deg(g)/2$ queries to the $n$ variables to compute $g$, and that
$O(deg(g)^4)$ queries suffice (see also~\cite{nisan&szegedy:degree}).
This implies that classical and quantum query complexity are 
polynomially related for all total $f$.
Similarly, we have shown here that $(\log mon(g))/2$ qubits need to be
communicated to compute $f(x,y)=g(x\wedge y)$.
An analogous upper bound like $Q^*(f)\in O((\log mon(g))^k)$ might be true.
A similar resemblance holds in the bounded-error case.
Let $\widetilde{deg}(g)$ be the minimum degree of polynomials that
approximate $g$. 
In~\cite{bbcmw:polynomials} it was shown that a bounded-error 
quantum computer needs at least $\widetilde{deg}(g)/2$ queries to
compute $g$ and that $O(\widetilde{deg}(g)^6)$ queries suffice.
Here we showed that $(\log\widetilde{m}(f))/2$ qubits of communication
are necessary to compute $f$.
A similar upper bound like $Q_2(f)\in O((\log\widetilde{m}(f))^k)$ may hold.

A second open question: 
how do we prove good lower bounds on {\em bounded-error} quantum protocols?
Theorems~\ref{thlogrankappr} and~\ref{thapprmon} of the previous section show 
that $Q_2(f)$ is lower bounded by $\log\widetilde{m}(f)/2$ and
$\log\widetilde{mon}(g)$ is lower bounded by $\sqrt{\so(g)}$.
If we could show $\widetilde{m}(f)\approx\widetilde{mon}(g)$ whenever 
$f(x,y)=g(x\wedge y)$, we would have $Q_2(f)\in\Omega(\sqrt{\so(g)})$.
Since $m(f)=mon(g)$ in the exact case, this may well be true.
As mentioned above, this is particularly interesting because it would
give a near-optimal lower bound $Q_2(\DISJ)\in\Omega(\sqrt{n})$.

Third and last, does prior entanglement add much power to qubit communication, or are
$Q(f)$ and $Q^*(f)$ roughly equal up to small additive or multiplicative
factors? Similarly, are $Q_2(f)$ and $Q^*_2(f)$ roughly equal?
The biggest gap that we know is $Q_2(\EQ)\in\Theta(\log n)$ versus
$Q^*_2(\EQ)\in O(1)$.

\bigskip

\noindent
{\bf Acknowledgments.}
We acknowledge helpful discussions with Alain Tapp, who first came up with 
the idea of reusing entanglement used in Section~\ref{seclogrank}. 
We also thank Michael Nielsen, Mario Szegedy, Barbara Terhal for discussions,
and John Tromp for help with the proof of Lemma~\ref{lemhypblset} 
in Appendix~\ref{appmonbound}.

\appendix
\section{Proof of Theorem~\ref{thlogrankclean}}\label{applogrankclean}

Here we prove a $\log rank(f)$ lower bound for clean qubit protocols.

\begin{lemma}[Kremer/Yao]\label{lemkremer}
The final state of an $\ell$-qubit protocol (without prior entanglement)
on input $(x,y)$ can be written as
$$
\sum_{i\in\01^\ell}\alpha_i(x)\beta_i(y)\st{A_i(x)}\st{i_\ell}\st{B_i(y)},
$$
where the $\alpha_i(x),\beta_i(y)$ are complex numbers and
the $A_i(x),B_i(y)$ are unit vectors.
\end{lemma}

\begin{proof}
The proof is by induction on $\ell$:

{\bf Base step.} 
For $\ell=0$ the lemma is obvious.

{\bf Induction step.} 
Suppose after $\ell$ qubits of communication the state can be written as
\begin{equation}\label{eqcommstate}
\sum_{i\in\01^\ell}\alpha_i(x)\beta_i(y)\st{A_i(x)}\st{i_\ell}\st{B_i(y)}.
\end{equation}
We assume without loss of generality that it is Alice's turn: 
she applies $U_{\ell+1}(x)$ to her part and the channel.
Note that there exist complex numbers $\alpha_{i0}(x),\alpha_{i1}(x)$ 
and unit vectors $A_{i0}(x),A_{i1}(x)$ such that
$$
(U_{\ell+1}(x)\otimes I)\st{A_i(x)}\st{i_\ell}\st{B_i(y)}=
\alpha_{i0}(x)\st{A_{i0}(x)}\st{0}\st{B_i(y)}+\alpha_{i1}(x)\st{A_{i1}(x)}\st{1}\st{B_i(y)}.
$$
Thus every element of the superposition (\ref{eqcommstate}) ``splits in two'' when we apply $U_{\ell+1}$.
Accordingly, we can write the state after $U_{\ell+1}$ in the form required by the lemma.
\end{proof}

\bigskip

\noindent
{\bf Theorem~\ref{thlogrankclean}}\
$\displaystyle Q_c(f)\geq\log rank(f)+1$.

\bigskip

\begin{proof}
Consider a clean $\ell$-qubit protocol for $f$. 
By Lemma~\ref{lemkremer}, we can write its final state as
$$
\sum_{i\in\01^\ell}\alpha_i(x)\beta_i(y)\st{A_i(x)}\st{i_\ell}\st{B_i(y)}.
$$
The protocol is clean, so the final state is $\st{0}\st{f(x,y)}\st{0}$.
Hence all parts of $\st{A_i(x)}$ and $\st{B_i(y)}$ other than $\st{0}$ will
cancel out, and we can assume without loss of
generality that $\st{A_i(x)}=\st{B_i(y)}=\st{0}$ for all $i$.
Now the amplitude of the $\st{0}\st{1}\st{0}$-state is simply the sum 
of the amplitudes $\alpha_i(x)\beta_i(y)$ of the $i$ for which $i_\ell=1$. 
This sum is either 0 or 1, and is the acceptance probability $P(x,y)$ of the
protocol.
Letting $\alpha(x)$ (resp.~$\beta(y)$) be the dimension-$2^{\ell-1}$ vector
whose entries are $\alpha_i(x)$ (resp.~$\beta_i(y)$) for the $i$ with
$i_\ell=1$:
$$
P(x,y)=\sum_{i:i_\ell=1}\alpha_i(x)\beta_i(y)=\alpha(x)^T\cdot\beta(y).
$$
Since the protocol is exact, we must have $P(x,y)=f(x,y)$.
Hence if we define $A$ as the $|X|\times d$ matrix having the $\alpha(x)$ as rows 
and $B$ as the $d\times|Y|$ matrix having the $\beta(y)$ as columns, then $M_f=AB$.
But now
$rank(M_f)=rank(AB)\leq rank(A)\leq d\leq 2^{l-1},$
and the theorem follows.
\end{proof}

\section{Proof of Theorem~\ref{thnlogrank}}\label{appsakslovasz}

\noindent
{\bf Theorem~\ref{thnlogrank} (Lov\'{a}sz and Saks)}\
$D(f)\leq (1+\log(C^1(f)+1))(2+\log rank(f))$.

\bigskip

\begin{proof}
We will first give a protocol based on a 0-cover.
Let $c=C^0(f)$ and $R_1,\ldots,R_c$ be an optimal 0-cover.
Let $R_i=S_i\times T_i$. We will also use $S_i$ to denote the $|S_i|\times 2^n$
matrix of $S_i$-rows and $T_i$ for the $2^n\times|T_i|$ matrix of $T_i$-columns.
Call $R_i$ {\em type~1} if $rank(S_i)\leq rank(M_f)/2$, and {\em type~2} otherwise.
Note that $rank(S_i)+rank(T_i)\leq rank(M_f)$,
hence at least one of $rank(S_i)$ and $rank(T_i)$ is $\leq rank(M_f)/2$.

The protocol is specified recursively as follows.
Alice checks if her $x$ occurs in some type~1 $R_i$.
If no, then she sends a 0 to Bob;
if yes, then she sends the index $i$ and they continue with the reduced 
function $g$ (obtained by shrinking Alice's domain to $S_i$), 
which has $rank(g)=rank(S_i)\leq rank(M_f)/2$.
If Bob receives a 0, he checks if his $y$ occurs in some type~2 $R_j$.
If no, then he knows that $(x,y)$ does not occur in any $R_i$, 
so $f(x,y)=1$ and he sends a 0 to Alice to tell her;
if yes, then he sends $j$ and they continue with the reduced function $g$, 
which has $rank(g)=rank(T_i)\leq rank(M_f)/2$ because $R_j$ is type~2. 
Thus Alice and Bob either learn $f(x,y)$ or reduce to a function $g$ 
with $rank(g)\leq rank(f)/2$, at a cost of at most $1+\log(c+1)$ bits. 
It now follows by induction on the rank that 
$D(f)\leq (1+\log(C^0(f)+1))(1+\log rank(f))$.

Noting that $C^1(f)=C^0(\overline{f})$ and $|rank(f)-rank(\overline{f})|\leq 1$,
we have $D(f)=D(\overline{f})\leq (1+\log(C^0(\overline{f})+1))(1+\log rank(\overline{f}))
\leq (1+\log(C^1(f)+1))(2+\log rank(f))$.
\end{proof}

\section{Proof of Theorem~\ref{thlogrankappr}}\label{applogrankappr}

\noindent
{\bf Theorem~\ref{thlogrankappr}}\
$\displaystyle Q_2(f)\geq\frac{\log\widetilde{m}(f)}{2}$.

\bigskip

\begin{proof}
By Lemma~\ref{lemkremer} we can write the final state of 
an $\ell$-qubit bounded-error protocol for $f$ as
$$
\sum_{i\in\01^\ell}\alpha_i(x)\beta_i(y)\st{A_i(x)}\st{i_\ell}\st{B_i(y)}.
$$
Let 
$\phi(x,y)=\sum_{i\in\01^{\ell-1}}\alpha_{i1}(x)\beta_{i1}(y)\st{A_{i1}(x)}\st{1}\st{B_{i1}(y)}$
be the part of the final state that corresponds to a 1-output of the protocol.
For $i,j\in\01^{\ell-1}$, define functions $a_{ij},b_{ij}$ by
$$
a_{ij}(x)=\overline{\alpha_{i1}(x)}\alpha_{j1}(x)\inp{A_{i1}(x)}{A_{j1}(x)}
$$ 
$$
b_{ij}(y)=\overline{\beta_{i1}(y)}\beta_{j1}(y)\inp{B_{i1}(y)}{B_{j1}(y)}
$$
Note that the acceptance probability is
$$
P(x,y)=\inp{\phi(x,y)}{\phi(x,y)}
% & = & 
%\sum_{i\in\01^{\ell-1}}\overline{\alpha_{i1}(x)\beta_{i1}(y)}\bra{A_{i1}(x)}\bra{1}\bra{B_{i1}(y)}
%\cdot\\
% & & \sum_{j\in\01^{\ell-1}}\alpha_{j1}(x)\beta_{j1}(y)\st{A_{j1}(x)}\st{1}\st{B_{j1}(y)}\\
=\sum_{i,j\in\01^{\ell-1}}a_{ij}(x)b_{ij}(y).
$$
We have now decomposed $P(x,y)$ into $2^{2\ell-2}$ functions.
However, we must have $|P(x,y)-f(x,y)|\leq 1/3$ for all $x,y$, 
hence $2^{2\ell-2}\geq \widetilde{m}(f)$.
It follows that $\ell\geq(\log\widetilde{m}(f))/2 +1$.
\end{proof}

\section{Proof of Theorem~\ref{thapprmon}}\label{appmonbound}

Here we prove Theorem~\ref{thapprmon}.
The proof uses some tools from the degree-lower bound proofs of Nisan 
and Szegedy~\cite[Section~3]{nisan&szegedy:degree}, including the following result
from~\cite{ehlich&zeller:schwankung,rivlin&cheney:approx}:

\begin{theorem}[Ehlich, Zeller; Rivlin, Cheney]\label{thehlichzeller}
Let $p$ be a single-variate polynomial of degree $deg(p)$ such that
$b_1\leq p(i)\leq b_2$ for every integer $0\leq i\leq n$, and
the derivative satisfies $|p'(x)|\geq c$ for some real $0\leq x\leq n$.
Then $deg(p)\geq\sqrt{cn/(c+b_2-b_1)}$.
\end{theorem}

A {\em hypergraph} is a set system $H\subseteq{\cal P}ow\{1,\ldots,n\}$.
The sets $E\in H$ are called the {\em edges} of $H$.
We call $H$ an {\em $s$-hypergraph} if all $E\in H$ satisfy $|E|\geq s$.
A set $S\subseteq\{1,\ldots,n\}$ is a {\em blocking set} for $H$ if
it ``hits'' every edge: $S\cap E\neq\emptyset$ for all $E\in H$.

\begin{lemma}
Let $g:\01^n\rightarrow\01$ be a Boolean function for which 
$g(\vec{0})=0$ and $g(e_i)=1$, $p$ be a multilinear polynomial which approximates $g$
(i.e.~$|g(x)-p(x)|\leq 1/3$ for all $x\in\01^n$),
and $H$ be the $\sqrt{n/12}$-hypergraph formed by the set of all monomials 
of $p$ that have degree $\geq\sqrt{n/12}$.
Then $H$ has no blocking set of size $\leq n/2$.
\end{lemma}

\begin{proof}
Assume, by way of contradiction, that there exists a blocking set $S$
of $H$ with $|S|\leq n/2$.
Obtain restrictions $h$ and $q$ of $g$ and $p$,
respectively, on $n-|S|\geq n/2$ variables by fixing all $S$-variables to 0.
Then $q$ approximates $h$ and all monomials of $q$ have degree
$<\sqrt{n/12}$ (all $p$-monomials of higher degree have been set to 0
because $S$ is a blocking set for $H$).
Since $q$ approximates $h$ we have $q(\vec{0})\in[-1/3,1/3]$,
$q(e_i)\in[2/3,4/3]$, and $q(x)\in[-1/3,4/3]$ for all other $x\in\01^n$.
By standard symmetrization techniques~\cite{minsky&papert:perceptrons,nisan&szegedy:degree}, 
we can turn $q$ into a single-variate polynomial $r$ of degree $<\sqrt{n/12}$,
such that $r(0)\in[-1/3,1/3]$, $r(1)\in[2/3,4/3]$, 
and $r(i)\in[-1/3,4/3]$ for $i\in\{2,\ldots,n/2\}$.
Since $r(0)\leq 1/3$ and $r(1)\geq 2/3$, we must have $p'(x)\geq 1/3$ for some real $x\in[0,1]$.
But then $deg(r)\geq\sqrt{(1/3)(n/2)/(1/3+4/3+1/3)}=\sqrt{n/12}$ 
by Theorem~\ref{thehlichzeller}, contradiction.
%Hence there is no blocking set $S$ with $|S|\leq n/2$.
\end{proof}

The next lemma shows that $H$ must be large if it has no
blocking set of size $\leq n/2$:

\begin{lemma}\label{lemhypblset}
If $H$ is an $s$-hypergraph of size $m<2^{s}$,
then $H$ has a blocking set of size $\leq n/2$.
\end{lemma}

\begin{proof}
We use the probabilistic method to show the existence of a blocking set $S$.
Randomly choose a set $S$ of $n/2$ elements.
The probability that $S$ does not hit some specific $E\in H$ is
%(using $1-x\leq e^{-x}$)
$$
\frac{{n-|E|\choose n/2}}{{n\choose n/2}}=
\frac{\frac{n}{2}(\frac{n}{2}-1)\ldots(\frac{n}{2}-|E|+1)}{n(n-1)\ldots(n-|E|+1)}\leq
%\frac{(n-|E|)(n-|E|-1)\ldots((1-b)n-|E|+1)}{n(n-1)\ldots((1-b)n+1)}\leq
%\left(1-\frac{|E|}{n}\right)^{bn}\leq
2^{-|E|}.
$$
Then the probability that there is some edge $E\in H$ which is not hit
by $S$ is
$$
\Pr[\bigvee_{E \in H} \mbox{ S does not hit E}]\leq
\sum_{E\in H} \Pr[\mbox{S does not hit E}]\leq
\sum_{E\in H} 2^{-|E|}\leq m\cdot 2^{-s}<1.
$$
Thus with positive probability, $S$ hits all $E\in H$,
which proves the existence of a blocking set.
% of size $n/2$.
\end{proof}

The above lemmas allow us to prove:

\bigskip

\noindent
{\bf Theorem~\ref{thapprmon}}\
{\it If $g$ is a Boolean function, then
$\displaystyle\widetilde{mon}(g)\geq 2^{\sqrt{\so(g)/12}}.$}

\bigskip

\begin{proof}
Let $p$ be a polynomial which approximates $g$ with 
$\widetilde{mon}(g)$ monomials.
Let $b=\so(g)$, and $z$ and $S_1,\ldots,S_b$ be the input and sets which
achieve the 0-block sensitivity of $g$.
We assume without loss of generality that $g(z)=0$.

We derive a $b$-variable Boolean function $h(y_1,\ldots,y_b)$ 
from $g(x_1,\ldots,x_n)$ as follows:
if $j\in S_i$ then we replace $x_j$ in $g$ by $y_i$, and 
if $j\not\in S_i$ for any $i$, then we fix $x_j$ in $g$ to the value $z_j$.
Note that $h$ satisfies
\begin{enumerate}
\item $h(\vec{0})=g(z)=0$
\item $h(e_i)=g(z^{S_i})=1$ for all unit $e_i\in\01^b$
\item $\widetilde{mon}(h)\leq\widetilde{mon}(g)$, because we can easily derive
an approximating polynomial for $h$ from $p$, without increasing the number
of monomials in $p$.
\end{enumerate}
It follows easily from combining the previous lemmas that any approximating 
polynomial for $h$ requires at least $2^{\sqrt{b/12}}$ monomials,
which concludes the proof.
\end{proof}

\end{document}